\documentclass[showpacs,amsmath,amssymb,twocolumn,prl, floatfix]{revtex4}
\usepackage[dvips]{graphicx}
\input{epsf}

\begin{document}

\title{Wave chaos in rapidly rotating stars}

\author
{Fran\c{c}ois Ligni\`eres$^{1}$}
\email{ligniere@ast.obs-mip.fr}
\author
{Bertrand Georgeot$^2$}
\affiliation{$^1$Laboratoire d'Astrophysique de Toulouse-Tarbes,
Universit\'e de Toulouse, CNRS,
31400 Toulouse,
France \\
$^2$Laboratoire de Physique Th\'eorique, Universit\'e de Toulouse, UPS, CNRS,
 31062 Toulouse,
France}

\date{\today}

\begin{abstract}
Effects of rapid stellar rotation on acoustic
oscillation modes are poorly understood. We study the dynamics of acoustic rays
in rotating polytropic stars
and show using quantum chaos concepts that
the eigenfrequency spectrum
is a superposition of regular frequency patterns and
an irregular frequency subset respectively associated with near-integrable and
chaotic phase space regions.
This opens new perspectives for rapidly rotating star seismology and also
provides a new and potentially observable manifestation of wave chaos
in a large scale natural system.
\end{abstract}
\pacs{97.10.Sj, 05.45.Mt, 97.10.Kc}
\maketitle

Since helioseismology revolutionized our knowledge of the Sun's interior, many efforts including space missions (MOST, COROT and KEPLER) are undertaken
to detect oscillation frequencies in a large variety of stars \cite{Chri,Saio}. 
But to access the information contained in these data, the observed frequencies must be first associated with the right 
stellar oscillation modes. This crucial identification process requires a full understanding 
of the properties of the oscillation spectrum and,
for slowly rotating stars like the Sun, the asymptotic theory of high frequency acoustic modes
provided such an understanding \cite{Chri}.
Both the approximate treatment of the centrifugal distortion \cite{Goup} and the lack
of asymptotic theory have so far hindered reliable identifications in rapidly rotating pulsators. This long-standing problem
mainly concerns massive and intermediate-mass stars \cite{Kurt} like the $\delta$ Scuti star Alta\"ir whose surface oblateness
has been measured by interferometry \cite{Alta}.
Accurate computations of acoustic modes fully taking into account the effect of rotation on the oscillations
have only recently been performed for polytropic models of rotating stars \cite{Lign}.
Here we construct
the dynamics of acoustic rays to understand 
the properties
of the frequency spectrum. 

The acoustic ray model 
is 
analogous to the geometrical optics limit of
electromagnetic waves or the
classical limit of quantum mechanics.
The construction of eigenmodes from stellar acoustic rays has been already considered 
in the integrable case of a non-rotating spherically symmetric star \cite{Goug}.
However when the ray dynamics is no longer integrable, the problem is known to become of deeply different nature.
This issue has been mostly investigated by the quantum chaos community 
in the context of the classical limit of
quantum systems \cite{Houc} and the developed concepts have been applied to other wave phenomena such as those
observed in e.g. microwave resonators \cite{Stoc},
lasing cavities \cite{Nock},
quartz blocks \cite{Elle}
and
underwater waves \cite{Brow}. The potential interest for stellar seismology has been suggested \cite{Perd} but not yet demonstrated.

Our star model is a self-gravitating uniformly rotating monoatomic perfect gas ($\Gamma=5/3$) where pressure and density follow
a polytropic relation $P_e \propto \rho_e^{1+1/N}$ with $N=3$.
Neglecting the Coriolis force and the gravitational potential perturbations, small amplitude adiabatic perturbations around this equilibrium 
verify:
\begin{eqnarray}
\!\!\!\!{\partial}_t \rho + {\bf \nabla}\!\!\cdot\!\!(\rho_{e} {\bf v})=0,  & {\partial}_t {\bf v} =
- \frac{{\bf \nabla} P}{\rho_{e}} + \frac{\rho}{\rho_{e}} {\bf {g}_e},  & dP = c_s^{2} d\rho
\label{first}
\end{eqnarray}
where the density $\rho$, pressure $P$ and velocity ${\bf v}$ describe the perturbation while $c_s$ is the sound velocity 
and $\bf {g}_e$ is the effective gravity resulting from
the gravitational and centrifugal potentials. As other quantities characterizing the star model, $c_s$ and $\bf {g}_e$
vary in the meridian plane of the rotating star.
Neglecting gravity waves, these equations can be reduced to the form, $({\omega}_c^2 - {\omega}^2) \Phi - c_s^2 \Delta \Phi=0$
where $\Phi=\hat{P}/c_s^{3}$ is related to the time-harmonic pressure perturbation $P=\Re\lbrace\hat{P} \exp (- i \omega t)\rbrace$
and ${\omega}_c = \sqrt{(15/64) ({\bf g}_{\bf e}/c_s)^2
+(3/8) {\bf \nabla}\!\cdot\!{\bf g_e}}$
is the cut-off frequency whose sharp increase in the outermost layers of the star provokes the
back reflection of acoustic waves. 
The WKB approximation then leads to the eikonal equation,
$\omega^2 = c_s^2 {\bf k}^2 + {\omega}_c^2 $.
The acoustic ray is the trajectory tangent to the wave vector 
${\bf k}$ at the point ${\bf x}$ and its evolution
can be described by Hamilton's equations, $H=\sqrt{c_s^2 {\bf k}^2 + {\omega}_c^2}$ being the Hamiltonian \cite{Goug}.
Rays heading towards the star center tends to be refracted by increasing sound velocity while close to the surface non-specular reflection
takes place at $\omega_c = \omega$. As rotation increases, isocontours of $c_s$ and $\omega_c$ are distorted together with the star surface.
In the following, we restrict ourselves to axisymmetric modes $L_z=0$, thus reducing the phase space to four dimensions.

\begin{figure*}[htb]
\begin{center}
\includegraphics[width=0.95\linewidth]{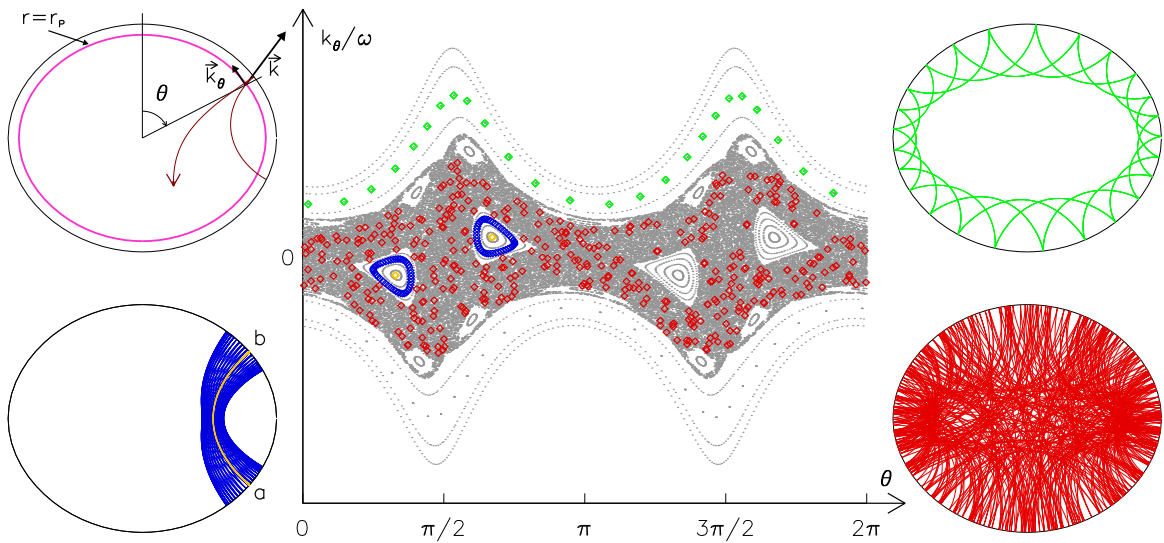}
\end{center}

\caption{(Color online) PSS and typical acoustic rays at a rotation equal to $59$ percent of the Keplerian limit. 
A whispering gallery ray (green/light grey), an island ray (blue/dark grey) and a chaotic ray (red/grey) 
are shown on the physical space and on the PSS (diamonds in the center figure). The central orbit of the island is also shown (yellow between points $a$ and $b$).
$k_{\theta}/\omega$ is in unit of $\sqrt{r_{s}^3 (0) / GM}$ with $M$ the stellar mass and $G$ the gravitational constant.
}
\end{figure*}

The acoustic ray dynamics has been investigated by integrating numerically the Hamilton's equations and the 
resulting dynamics is visualized using the standard tool of Poincar\'e surface of section (PSS). 
As rays do not reach the star boundary, the PSS is
defined by $r_p (\theta)=r_s (\theta) - d$ where the distance $d=0.08 r_s (\pi/2)$ from the star surface $r_s(\theta)$
has been chosen such that all but a few whispering gallery rays cross the PSS (only outgoing rays are taken).
The two coordinates of the PSS are
$\theta$, the colatitude, and $k_{\theta}/\omega$, $k_{\theta}$ being the angular component
of ${\bf k}$
in the
natural basis associated with the coordinate system $(\zeta=r_s (\theta)-r, \theta)$.
We use the scaled variable $k_{\theta}/\omega$
as, in the limit $\omega \gg {\omega}_c$, the ray dynamics
becomes independent of the frequency away from the reflection points.
We found that increasing the stellar rotation leads to a soft transition from integrability to chaos analogous to the
one described by the KAM theorem.
As illustrated in Fig. 1 for a given rotation rate, 
the phase space shows a mixed structure where chaotic regions coexist with a whispering gallery region 
close to the boundary
and regular islands around stable periodic orbits.
As rotation increases, both the chaotic region and the central island chain get larger.
A crucial feature of the dynamics is that
each region is dynamically isolated from the other by invariant tori which prevent communication between them.
Such a situation has been found several times in the domain 
of quantum chaos, and generally it was surmised \cite{Berr}
that the stationary waves localized on one of these regions 
form an independent subset with specific dynamical properties.
The frequency spectrum thus appears as the superposition of independent frequency subsets 
reflecting the phase space structure. This surmise has been found to be a good approximation for many systems,
although some correlations may remain between the frequency subsets
due to modes localized at the border between zones or due
to the presence of partial barriers in phase space \cite{Bohia}.

\begin{figure}[htb]
\begin{center}
\includegraphics[width=0.95\linewidth]{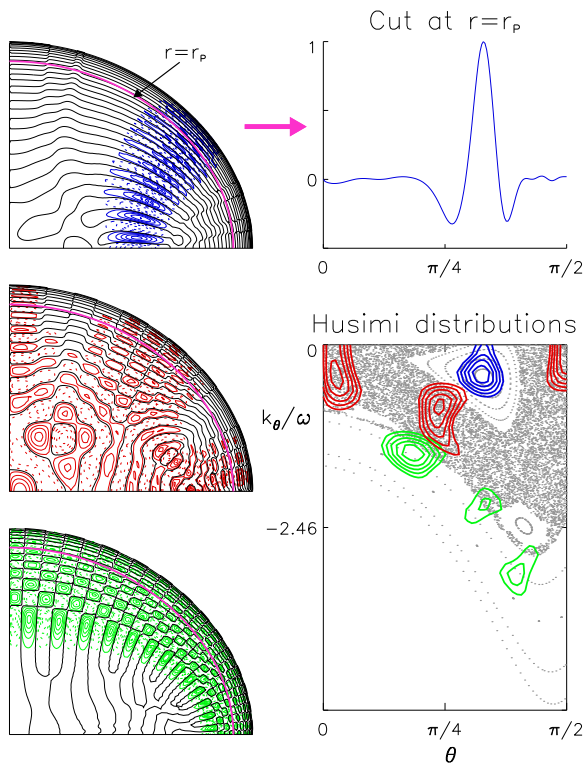}
\end{center}

\caption{(Color online) Comparison of numerically computed acoustic modes from (\ref{first}) with phase space ray dynamics on the PSS.
Three typical modes have been selected: island mode (blue/dark grey), chaotic mode (red/grey), whispering gallery
mode (green/light grey). Left: spatial distribution of $\Phi'$
on one quarter of a meridian plane.
Black lines are the nodal lines, full (resp. dashed) colored/grey lines are
the level curves for positive (resp. negative) values. Magenta line at $r=r_p$ is the chosen PSS.
Right top: cut of the island mode along the PSS, from the pole to the equator. Right bottom:
level curves of the Husimi distributions of the three modes (for the same of value of $\Delta=0.12 r_s(\pi/2 )$),
showing that they are mainly concentrated inside the main stable island (for the island mode), in the chaotic region
(for the chaotic mode), or in the whispering gallery region (for the whispering gallery mode).}
\end{figure}

It is therefore important to know if
this spectrum organization is valid in the high frequency limit where
it is supposed to hold, and even more important to assess if
it is still relevant to the observable
acoustic modes. We have thus numerically computed exact axisymmetric modes of Eq. (\ref{first}), using the method described in \cite{Lign},
in the frequency range $[{\omega}_1 , 12 {\omega}_1 ]$,
$\omega_1$ being the lowest acoustic frequency.
To establish a link with
the asymptotic ray dynamics, we use
a phase-space representation of the modes known as Husimi distribution \cite{Chan}. In order to compare
a three-dimensional mode with the
acoustic rays on a two-dimensional plane, the mode amplitude is first scaled by the square root of the distance to the rotation axis \cite{Goug}.
The Husimi distribution is then
constructed from a cut taken along the PSS:
$H(s_0,k_{0})= |\int \Phi'(s) \exp (-(s-s_0)^2/(2\Delta^2)) \exp (ik_0 s) ds|^2$.
Here $\Phi'$ is the scaled version of $\Phi$, the integral is taken along the curve $r=r_p$ and $k$ is the moment in the direction tangent
to this curve; $\Delta$ is the width of the Gaussian wavepacket on which $\Phi'$ is projected. In Fig. 2,
$(s_0,k_{0})$ is replaced by $(\theta,k_{\theta})$ for comparison with data from Fig. 1. 
Except for some of the largest lengthscale modes close to the frequency $\omega_1$, 
we find that the Husimi distribution enables to unambiguously 
associate the modes with the main structures of phase space. 
As illustrated in Fig. 2, we distinguished the island modes trapped into the main stable islands, the chaotic modes localized in
the central chaotic
region and the whispering gallery modes associated with the whispering gallery region.
We note that, due to the relatively low frequency considered, the chaotic modes do not spread over all parts of the chaotic region.

\begin{figure}[htb]
\begin{center}
\includegraphics[width=.95\linewidth]{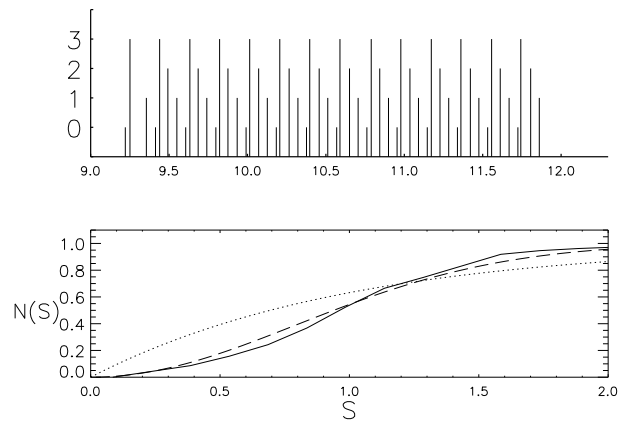}
\end{center}

\caption{
Top: island mode frequencies in the interval $[9 \omega_1, 12 \omega_1]$ showing the regular spacings corresponding
to formula (\ref{numbers}); height reflects the value of the
quantum number $\ell$.  Bottom: 
Integrated spacing distribution for the chaotic modes in the same interval (the full line).
Data correspond to around $200$ modes from two symmetry classes.
The dashed line is the result for the Gaussian Orthogonal Ensemble of Random Matrix Theory while
the dotted line corresponds to the Poisson
distribution characteristic of integrable systems.
}
\end{figure}

Having defined subsets of modes, we can now analyze the properties of the corresponding frequency subsets.
As shown in Fig. 3, the frequency spacings of the island modes display striking regularities which lead to the simple
empirical formula:
\begin{equation}
\omega_{n \ell} = n {\delta}_n + {\ell} {\delta}_{\ell} + \alpha
\label{numbers}
\end{equation}
where $n$ and ${\ell}$ are natural integers, ${\delta}_n$ and ${\delta}_{\ell}$ are uniform frequency spacings. 
The $\alpha$ constant being fixed by a given island mode frequency, the formula proves sufficiently accurate to
identify the other island modes among the whole set of computed frequencies.
The phase space representation of these modes reveal that these regular patterns can be attributed 
to the existence of the stable island region in phase space.
Although a zoom on this region would show a complex structure involving chaotic trajectories and chains 
of small islands nested between deformed surviving 
tori, these small scale details can be overlooked for the relatively large wavelengths considered here.
To retrieve formula (\ref{numbers}) and to find out how ${\delta}_n$ and ${\delta}_{\ell}$ relate to the 
properties of the star, we follow an approach inspired by the quantization of laser modes in cavities \cite{Koge}.
Indeed, far from the boundary, our problem can be translated into the propagation of light in an inhomogeneous
medium, $1/c_s$ playing the role of the medium index. Close to the stable orbit, we can apply the paraxial approximation.
In this case, it is known that the wave beam solution is \cite{Perm} $\Phi(\sigma,\xi) \propto H_{\ell}(\sqrt{2}\xi/w(\sigma))
\exp(-\xi^2/w(\sigma)^2)
\exp(-i \phi(\sigma,\xi))$, where $\sigma,\xi$ are coordinates parallel and transverse
to the central periodic orbit (the yellow curve in Fig. 1), $H_{\ell}$ is the Hermite polynomial of degree $\ell$.
The spreading of the beam in the transverse direction is described by $w(\sigma)$ which verifies
$(1/c_s) d/d\sigma [(1/c_s)(dw/d\sigma)] +
\alpha(\sigma) w=4/w^3$, where $\alpha(\sigma)=(1/c_s^3) \partial^2 c_s/\partial \xi^2$. The wave phase
is $\phi(\sigma,\xi)) = \omega \int_{0}^{\sigma} d\sigma' /c_s -2({\ell}+1) \int_{0}^{\sigma} c_s d\sigma' /w^2 + \xi^2/(2c_s R)$
where $R=w/(d w /d\sigma)$ is the radius of curvature of the beam wavefront.
Numerically computed island modes have a transverse variation confirming this approximation.
We then obtain a stationary solution by imposing that the wave interferes constructively with itself.
This requires that the phase accumulated following the periodic orbit ($\xi = 0$)
from one side of the boundary
to the other side is
$\omega \int_{\mbox{a}}^{\mbox{b}} d\sigma /c_s
-2({\ell}+1) \int_{\mbox{a}}^{\mbox{b}} c_s d\sigma /w^2 = n \pi$.
This 
leads to the formula (\ref{numbers})
with ${\delta}_n= \pi /(\int_{\mbox{a}}^{\mbox{b}} d\sigma/c_s)$
and ${\delta}_{\ell}= 
2 (\int_{\mbox{a}}^{\mbox{b}} c_s d\sigma /w^2)/(\int_{\mbox{a}}^{\mbox{b}} d\sigma/c_s)$. 
The numerical value of ${\delta}_n$ obtained from the island mode frequencies shown on Fig. 3 (equal to $0.5514$
in units of $\sqrt{G M/r_s^3 (0)}$ where
$M$ is the stellar mass and $G$ the gravitational constant) 
is well approximated, within 2.2 percent, by the theoretical one 
(equal to $0.5635$ in the same units).
While ${\delta}_n$ probes the sound velocity along the path of the periodic orbit,
${\delta}_{\ell}$ is obtained by solving the second order equation verified by $w$ together with
the two boundary conditions given by the necessity to match $R$ with
the radius of curvature of the two bounding surfaces.
Thus, ${\delta}_{\ell}$ probes the second order transverse derivative of the sound velocity along the same path
as well as the radius of curvature of the bounding surfaces.
We note that similar modes around stable periodic orbit have been constructed in other systems, usually
with the more systematic procedure of finding the normal forms and using EBK quantization \cite{Bohia, Lazu}.

Having shown that modes whose Husimi distribution is localized in the near-integrable region
display integrable-like quantization conditions, we now turn to the modes localized in the chaotic region.
The subset of chaotic mode frequencies shows typical signatures of wave chaos such as 
frequency repulsion. Indeed, in Fig. 3, the integrated distribution of
consecutive frequency spacings $S_i = \omega_{i+1} - \omega_{i}$ (normalized by the mean frequency spacing
of those modes)
is much closer to the Random Matrix Theory result typical of chaotic systems \cite{Bohib}
than to the Poisson distribution result characteristic of integrable systems.
This frequency statistics together with the fact that the corresponding modes are all localized 
in the chaotic region of the ray dynamics give a strong evidence that wave chaos occurs in rapidly rotating stars.
The difficulty to solve Eq. (\ref{first}) even with state of the art computational techniques prevents
us to reach a larger frequency sample and to make detailed comparison with Random Matrix Theory
as in e.g. \cite{Bohib}. 

The whispering gallery modes and the modes trapped in smaller island chains being associated with near-integrable
regions of phase space, their frequencies are therefore expected to display regular patterns. The detail study 
of these regularities shall be considered elsewhere as it requires more mode calculations with a higher numerical resolution.
It is also important to point out that these modes will be the most difficult to detect.
Indeed, due to their
small latitudinal wavelength (see Fig. 2), the positive and negative light intensity fluctuations
strongly cancel out when integrated over the visible disk.

Our results demonstrate that ray dynamics and quantum chaos concepts provide a qualitative and quantitative 
insight into the frequency spectrum of rapidly rotating stars.
In particular we are able
to separate the spectrum of a reasonably realistic star model into integrable and chaotic subsets.
Being much less demanding than the direct eigenmodes computation as well as easily adaptable to non-axisymmetric modes
(for which regular frequency patterns have also been found numerically \cite{Reesb})
and to more realistic stellar models, ray dynamics will be essential to further
specify the asymptotic properties of the oscillation spectrum.

The present analysis opens new perspectives in seismology of rapidly rotating stars.
Observed spectra differ from theoretical ones as
poorly understood mechanisms governing the intrinsic mode amplitude determine the frequencies that
are actually detected.
In this context, a priori information on the structure
of the spectrum is crucial in order to 
identify the observed frequencies with specific stellar oscillation modes.
Our results strongly suggest to first look for regular patterns to identify
the island modes and to determine seismic observables as ${\delta}_n$ and ${\delta}_{\ell}$ containing 
information on the star's interior.
The remaining chaotic modes are also of special interest for seismology purposes: they
are highly sensitive to small changes of the stellar model \cite{Scha} and, contrary to non-radial
acoustic modes of slowly rotating stars which avoid the star's center, they probe this
region which is crucial for stellar evolution theory. 
If enough chaotic modes are seen, their mean frequency spacing which is known to depend on the volume of the chaotic region
would constrain the stellar rotation.

We thank S. Vidal, D. Reese, M. Rieutord and L. Valdettaro for their help at
various stages of this work.
We also thank CALMIP ("CALcul en MIdi-Pyr\'en\'ees'') for the use of their supercomputer.

\end{document}